%% file: main.tex
\documentclass[12pt,a4paper]{article}
\usepackage[pdfstartview=FitH,colorlinks=true,linkcolor=black,anchorcolor=black,citecolor=black,urlcolor=black]{hyperref}
\usepackage[english]{babel}
\usepackage{amsmath,amssymb,titling,authblk}
\usepackage{slashed}
\usepackage{amsmath}
\usepackage{amscd}
\usepackage{mathrsfs}
\usepackage[normalem]{ulem}
\usepackage{appendix}
\usepackage{cite}

\usepackage{physics}
\usepackage{amsfonts,mathtools}

\ifx\pdfoutput\undefined
\usepackage{graphicx}
\else
\usepackage[pdftex]{graphicx}
\fi

\usepackage[font=small]{caption}
\usepackage{subcaption}
\usepackage{xcolor}

\usepackage{xcolor}
\usepackage{hyperref}
\hypersetup{pdfstartview=FitH, linkcolor=black,urlcolor=black, colorlinks=true}

\DeclareMathAlphabet{\mathpzc}{OT1}{pzc}{m}{it}

\begin{document}

\author[1]{Oleg O.~Novikov\thanks{o.novikov@spbu.ru}}

\author[1]{Andrey A.~Shavrin\thanks{shavrin.andrey.cp@gmail.com}}

\affil[1]{Saint Petersburg State University, 7/9 Universitetskaya emb.\\
St. Petersburg, 199034, Russia}

\title{Large D charged black hole as a Jackiw-Teitelboim gravity weakly coupled to the thermal graviton background}

\maketitle

\begin{abstract}
    The s-wave approximation to the black hole dynamics has attracted considerable attention recently. However, the near-AdS2 geometry of the near-horizon region and decoupling of the non-s-wave modes usually requires a small-temperature limit. In this work we propose the new limit based on the large number of dimensions in which the temperature remains finite, but the s-wave sector may be described by Jackiw-Teitelboim gravity interacting with other modes with vanishingly small coupling. However, we demonstrate that the thermal bath of the tensor modes with non-zero angular momenta collectively produces a significant impact on the two-dimensional theory.
\end{abstract}

\input{Introduction}

\input{NearExtremal}
\input{S-Wave}

\input{TensorCorrections}
\input{Termalization}

\input{Conclusion}

\bibliographystyle{habbrv}
\bibliography{ref}

\end{document}

%% file: Introduction.tex
\section{Introduction}

Black holes near-extremal solutions exhibit universal low-energy dynamics that are often captured by an effective two-dimensional description. In particular, the dimensional reduction yields a nearly-AdS${}_2$ in the near zone, that can be described in leading order by Jackiw–Teitelboim (JT) two-dimensional dilaton gravity. The s-wave perturbations of the $D$-dimensional theory correspond in this description to the dilaton~\cite{Navarro:1999qy,Fabbri:2000xh,Kraus:2025efu,Nayak:2018qej,Castro:2021fhc,Castro:2025itb}. This provides a degree of freedom to the otherwise purely topological theory of gravity in 2D.

JT gravity being the simplest realization of gravity in two-dimensions is a quantum integrable model~\cite{Saad:2019lba}. This makes its appearance in the description of the black hole highly interesting as it provides exact quantum theoretical results for the dynamics in this approximation. This may be compared with the description of the black-hole thermodynamics of near-extremal black holes, for which the s-wave approximation also plays the main role~\cite{Kraus:2025efu}.

From the semiclassical considerations, the extremal black hole should have zero temperature and emit no Hawking radiation. This serves as a first approximation to the low temperature black holes described by the near-extremal solution that allows one to study such aspects of the quantum behavior of the black holes as the existence of the mass gap in the spectrum.

The s-wave approximation prohibits one from using the microscopic description for the tensor and vector particles, as they must have transverse polarizations. This limits one to consider the averaged hydrodynamical description of the matter or scalar and spinor fields. That makes the correct microscopic account for the Hawking radiation impossible. Also to employ the s-wave approximation one must ensure that the modes with higher angular momentum decouple from the considered spherically symmetric sector. This may be achieved in the aforementioned low-temperature limit, however makes applications of the s-wave approximation to the finite temperature black holes questionable.

A different limit, that we employ in this work, is a large $D$ limit, where $D - 1$ denotes a number of spatial dimensions. For fixed radius of the black hole, its gravity is concentrated in the near horizon zone, whereas it can be neglected in the far zone\cite{Guo:2015swu,Emparan:2014cia,Emparan:2020inr}. A sharp distinction between the near horizon zone and far zone results in the dramatic simplification of the equations of motion (EoM). It is general behavior for all non-s-wave modes to interact only weakly with black hole. This can be associated with decreasing surface area of black hole $r_+^n \text{Vol} S^n$ (where $n=D-2$) in this limit~\cite{Emparan:2014aba}.

While s-wave approximation was employed before for the description of the low-temperature black holes in the large $D$ limit, in this work we demonstrate that with an appropriate rescaling one can combine the near-extremal behavior of the solution with the finite temperature. Combined with the decoupling of the non-s-wave modes this provides an interesting case where the black hole itself can be described as a 2D JT gravity weakly coupled to the thermal bath of the inhomogeneous modes in the approximately Minkowski spacetime.

Nevertheless, while the individual modes may be decoupled from the s-wave sector, their combined impact may still be important proper account of the thermalization of the black hole.
In this work we consider tensor perturbations of the near-extremal RN solution, that can be interpreted as gravitational waves scattering on black hole. After dimensional reduction they can be treated as additional scalar matter fields in terms of two-dimensional theory on the background of JT gravity, producing a source term on the boundary of the near-AdS${}_2$ spacetime. We quantize canonically the tensor modes and compute the contribution of their thermal bath to this source term.

In this work we do not consider gravi-photons since they would require a more detailed analysis of the perturbations of the linearized gravity. All types of the perturbations are coupled with the gauge filed obtained from Kaluza-Klein reduction, that dramatically increase the complexity of the problem. However, this generalization is widely understood within s-wave approximation~\cite{Iliesiu:2020qvm,Banerjee:2021vjy,Button:2013rfa,Moitra:2018jqs}.

In the next section \ref{Section:NearExtremal} we apply the large D limit to the near-extremal RN solution and demonstrate that an asymptotic AdS${}_2$ patch in the near-horizon zone can be obtained at finite temperature.
Section \ref{Section:DimensionalReduction} is devoted to the construction of the effective two-dimensional JT gravity through a dimensional reduction of the $D$-dimensional theory in the large $D$ limit.
In section \ref{Section:Corrections} we consider the tensor perturbations on the Reissner-Nordstr\"{o}m background, provide the solution in the large $D$ limit, as well as construct the effective interaction term of the ingoing tensor modes with the boundary of the two-dimensional asymptotic AdS${2}_2$ patch.
The final section \ref{Section:Quantization} is dedicated to quantization of the tensor perturbations in the thermal bath at finite temperature.

%% file: NearExtremal.tex
\section{Near-extremal Reissner-Nordstr\"{o}m solution in large D limit}\label{Section:NearExtremal}

We start with Einstein-Maxwell action in $D$-dimensional spacetime $\mathcal{M}_D$
with Gibbons-Hawking-York term on the boundary $\partial\mathcal{M}_D$
\begin{multline}\label{NearExtremal:EM-action}
    I_{\mathcal{M}_D}
    = \frac{M_{Pl}^n}{2}\int_{\mathcal{M}_D} d^{n+2}X \sqrt{-g_{\mathcal{M}_D}} R_{\mathcal{M}_D}
    - \frac{1}{4} \int_{\mathcal{M}_D} d^{n+2}X \sqrt{-g_{\mathcal{M}_D}} F^2
    \\
    + M_{Pl}^n \int_{\partial\mathcal{M}_D} d^{n+1}y \sqrt{-h_{\partial\mathcal{M}_2}} K_{\mathcal{M}_D},
\end{multline}
here $D = n + 2$, $R_{\mathcal{M}_D}$ is Ricci scalar, $F$ is the electromagnetic field tensor, $K_{\mathcal{M}_D}$ is extrinsic curvature and $h_{\partial\mathcal{M}_2} = \det h_{MN}^{\partial\mathcal{M}_2}$ is the determinant of the induced metric on the boundary $\partial\mathcal{M}_D$.

In the absence of the $D$-dimensional cosmological cosntant, the charged black hole is described by the Reisner-Nordstr\"{o}m (RN) solution,
\begin{equation}
    ds^2_{RN} = - f(r) dt^2 + \frac{dr^2}{f(r)} + e^{2v(r)} d\Omega_n^2, \quad
    f(r) = 1 - \frac{2r_s}{r^k} + \frac{r_Q^2}{r^{2k}}, \quad k = n - 1 = D - 3
\end{equation}
where $d\Omega_n$ is the volume element of the $S^n$ sphere, $r_s$ is the determined by the mass of the black hole, and $r_q$ by its electrical charge $q$. We keep $v(r) = \log r$ for generality. The electromagnetic strength is
\begin{equation}
    F = q r_+^n e^{-nv} dt \wedge dr,\quad r_Q^2 = \frac{q^2}{n(n+1)M_{Pl}^n}.
    \label{ElectromagneticField}
\end{equation}

The blackening function of the RN solution can be represented as,
\begin{equation}
    f(r) = \left(1 - \frac{r_+^k}{r^k}\right)\left(1 - \frac{r_-^k}{r^k}\right), \quad
    r_\pm^k = r_s^k \pm \sqrt{r_s^{2k} - r_Q^{2k}}, \quad r_s \geq |r_q| 
\end{equation}
with $r_+$ outer and $r_-$ inner horizons.

The Hawking temperature of the black hole can be derived from the surface gravity, that is given as following in the case of diagonal metric
\begin{equation}
    T = \frac{|f'(r)|\big|_{r = r_+}}{4\pi}.
\end{equation}
The extremal solution corresponds to the limit when both the horizons match, that corresponds to zero temperature. The deviation from the extremal solution can be parametrized with a small parameter $\epsilon \ll 1$ as a difference between the inner and outer horizons
\begin{equation}
    r_+^k - r_-^k = \epsilon r_+^k, \quad
    \epsilon = \frac{\sqrt{r_s^{2k} - r_Q^{2k}}}{r_+^k}.
\end{equation}
The blackening function and the temperature can be rewritten in this terms,
\begin{equation}\label{NearExtremal:NearExtremalTemperature}
    f(r) = \left(1 - \frac{r_+^k}{r^k}\right)\left(1 - \frac{r_+^k}{r^k} + \epsilon \frac{r_+^k}{r^k}\right), \quad
    T = \frac{k}{4\pi} \frac{\epsilon}{r_+} + \mathcal{O}(\epsilon^2).
\end{equation}
The large D limit implies $k \to \infty$. We also keep the temperature finite and fixed assuming
\begin{equation}
    \epsilon = \frac{\tilde{\epsilon}}{k} \rightarrow 0, \quad
    T = \mathrm{fix}.
    \label{LargeDLimit}
\end{equation}
Below we show, that $\mathcal{M}_D$ can be split onto $\mathcal{M}_2 \times S^n$ in the near horizon zone and $\mathcal{M}_2$ is asymptotic AdS${}_2$.
The rescaled coordinate
\begin{equation}
    \varrho = k\Big(1-\frac{r_+^k}{r^k}\Big),
\end{equation}
whose finite values corresponds to the near horizon zone, provides the metric
\begin{equation}
    ds^2_{\mathcal{M}_D} =
    \frac{r_+^2}{k^2}\Big(
    - \varrho (\varrho + \tilde{\epsilon}) dt^2
    + \frac{d\varrho^2}{\varrho (\varrho + \tilde{\epsilon})}\Big)
    + r_+^2 d\Omega_n.
\label{RindlerAdSMetric}
\end{equation}
The following coordinate transformation
\begin{equation}
T 
= \frac{\exp(\tilde{\epsilon} t)}{\tilde{\epsilon}} \Big(\frac{\rho^2 + (\tilde{\epsilon} + \rho)^2}{\rho(\tilde{\epsilon} + \rho)}\Big) - 2\tilde{\epsilon}, \quad
Z
= \frac{\exp(\tilde{\epsilon} t)}{\tilde{\epsilon}} \Big(\frac{-\rho^2 + (\tilde{\epsilon} + \rho)^2}{\rho(\tilde{\epsilon} + \rho)}\Big)
\label{RindlerPatchTransform}
\end{equation}
transforms this metric into the metric of the Poincare patch of AdS${}_2\times S^n$ spacetime,
\begin{equation}
    ds^2_{\mathcal{M}_D} =
    \frac{r_+^2}{k^2}\Big(
    - \frac{dT^2}{Z^2}+\frac{dZ^2}{Z^2}\Big)
    + r_+^2 d\Omega_n
\end{equation}
An observation which will be useful for us below is that,
\begin{equation}
\frac{d}{dt}\frac{T + 2\tilde{\epsilon}}{Z}=0.
\label{RindlerTransformObservation}
\end{equation}

Note however that the original coordinates do not cover the whole Poincare patch but rather a part of it bounded by a lightlike line starting at $T=-2\tilde{\epsilon}, Z = 0$ and the Poincare patch lightlike asymptotic future $T\rightarrow+\infty$, representing past and future event horizons of the black hole correspondingly. This patch is commonly known as a Rindler patch, our transformation \eqref{RindlerPatchTransform} has the advantage that in the limit $\tilde{\epsilon}\rightarrow 0$ we smoothly transit to a Poincare patch which is a well known description of the near zone for the extremal black hole.

As the Rindler patch covers only a part of the AdS Cauchy surface, it may be expected that the quantum fields in it are in the thermal state. The point is that transitioning to \eqref{RindlerAdSMetric} does not imply rescaling the time. Therefore, this temperature should match the temperature of the black hole $T$.

Thus we have demonstrated that even for the finite $T$ the limit \eqref{LargeDLimit} results in the near-horizon zone being described by a patch of the AdS${}_2\times S_n$ spacetime.

%% file: S-Wave.tex
\section{Dimensional reduction in s-wave approximation}
\label{Section:DimensionalReduction}
To obtain an effective two-dimensional theory admitting near-AdS${}_2$, we employ the following convenient metric ansatz,
\begin{equation}
    ds_{\mathcal{M}_D}^2 = g^{\mathcal{M}_D}_{MN} \, dX^M\,dX^N = ds_{\mathcal{M}_2}^2 + r_{+}^2 e^{2v(x)} d\Omega_n^2,
\end{equation}
where $d\Omega_n^2$ is the metric on the unit sphere $S^n$ and the two-dimensional part is,
\begin{equation}
    ds_{\mathcal{M}_2}^2 = g^{\mathcal{M}_2}_{\mu\nu} dx^\mu dx^\nu,\quad
    \mu,\nu=0,1.
\end{equation}
The Ricci scalar from the Einstein-Hilbert (EH) term of the action in \eqref{NearExtremal:EM-action} has the dimensional reduction on the sphere $S^n$
\begin{equation}
    R_{\mathcal{M}_D} = R_{\mathcal{M}_2}
    + r_{+}^{-2} e^{-2v} R_{S^n}
    - 2n \nabla_{\mathcal{M}_2}^2 v
    - n(n+1)(\nabla v)^2_{\mathcal{M}_2},
\end{equation}
here $R_{\mathcal{M}_2}$ is Ricci scalar in $\mathcal{M}_2$, $R_{S^n} = n (n - 1)$ is Ricci scalar of the sphere.

The EH term contributes to the boundary terms of the two-dimensional theory with the dilaton field $v(x)$ due to Stokes theorem. Let's parametrize the boundary $\partial\mathcal{M}_2$ with the coordinate $u$ and embedding functions $(t(u),r(u))$. The timelike boundary has the tangent vector $\mathtt{t}^\mu$ with the Greek indexes $\mu,\nu,\alpha,\ldots$ in $\mathcal{M}_2$
\begin{equation}
    \mathtt{t}^\mu =\dot x^\mu = \begin{pmatrix}\dot t\\\dot r\end{pmatrix}, \quad
    \mathtt{t}_\mu\mathtt{t}^\mu = - 1,
\end{equation}
here dot $\dot{} = \tfrac{d}{du}$ denotes the derivative along the boundary coordinate. Later in the text we also use the uplifted vector $\mathtt{t}^M = (\mathtt{t}^t,\mathtt{t}^r,0,\ldots, 0)^\top \in T\mathcal{M}_D$ with the capital Latin indexes $M,N,A,B,\ldots$ in $\mathcal{M}_D$. The spacelike normal vector is
\begin{equation}
    \mathtt{n}_\mu = \sqrt{ - g_{tt} g_{rr}} \varepsilon_{\mu\nu} \mathtt{t}^\nu, \quad
    \varepsilon_{tr} = 1,
\end{equation}
that can be also trivially uplifted $\mathtt{n}^M = (\mathtt{n}^t,\mathtt{n}^r,0,\ldots, 0)^\top \in T\mathcal{M}_D$ due to the condition $n^M \perp S^n$.

One can define the induced metric on $\partial\mathcal{M}_2 \times S^n$ with the coordinates $y^{\hat a}$ with hatted Latin indexes $\hat a, \hat b, \ldots$ on $\partial\mathcal{M}_D$
\begin{equation}
    h_{\hat a \hat b} = g_{AB} e_{\hat a}^A e_{\hat b}^B, \quad
    h_{uu} = g_{\mu\nu} \mathtt{t}^\mu \mathtt{t}^\nu = - 1, \quad
    h_{ui}=0, \quad
    h_{ij} = g_{ij} = e^{2v} \gamma_{ij},
\end{equation}
here $e_{\hat a}^A: T\partial\mathcal{M}_D \to T\mathcal{M}_D$ are the vielbeins: $e_u^A = \mathtt{t}^A$, $e_i^A = (0,0,\delta_i^A)$ and $\gamma_{ij}$ is the metric with Latin indexes $i,j,k,l, \ldots$ on $S^n$.

The dimensionally reduced EH term takes the form
\begin{multline}
    I_{\mathcal{M}_2}^\text{EH}
    = \text{Vol} S^n \frac{(M_{Pl}r_{+}^n)}{2} \int_{\mathcal{M}_2} d^2x \sqrt{-g_{\mathcal{M}_2}} e^{nv} \left(
        R_{\mathcal{M}_2}
        + \frac{n (n - 1)}{r_{+}^2} e^{-2v}
        + n (n - 1) (\nabla v)^2_{\mathcal{M}_2}
    \right)
    \\ + \text{Vol} S^n (M_{Pl}r_{+})^n\int_{\partial\mathcal{M}_2} du \sqrt{-h} e^{nv} K_{\mathcal{M}_2}
\end{multline}
where $K_{\mathcal{M}_2}$ is the extrinsic curvature of $\partial\mathcal{M}_2$ as a wordline in $\mathcal{M}_2$. It is related  to the extrinsic curvature $K_{\mathcal{M}_D}$ in the GHY term of the action \eqref{NearExtremal:EM-action} as,
\begin{equation}
    K_{\mathcal{M}_D} = K_{\mathcal{M}_2} + n \mathtt{n}^\mu \nabla_\mu v, \quad
    K_{\mathcal{M}_2} = \nabla_\mu \mathtt{n}^\mu,
\end{equation}
where the extra term $n \mathtt{n}^\mu \nabla_\mu v$ is associated with the sphere $S^n$ of the radius $e^{v(r)}$.

The dimensionless coupling constant appearance in the following way
\begin{equation}
    \lambda \equiv \text{Vol} S^n (M_{Pl}r_{+})^n.
\end{equation}

The full dimensional reduction of the action \eqref{NearExtremal:EM-action} with the on-shell Maxwell term on the RN solution and electromagnetic field strength \eqref{ElectromagneticField} is
\begin{multline}
    I_{\mathcal{M}_2} =
    \frac{\lambda}{2}\int_{\mathcal{M}_2} d^2x \sqrt{-g_{\mathcal{M}_2}} e^{nv} \left(
        R_{\mathcal{M}_2}
        + \frac{n (n - 1)}{r_+^2} \Big(e^{-2v}-\frac{r_+^2}{r_Q^2}e^{-2nv}\Big)
        + n (n - 1) (\nabla v)^2_{\mathcal{M}_2}
    \right)
    \\ + \lambda
    \int_{\partial\mathcal{M}_2} du \sqrt{-h_{\partial\mathcal{M}_2}} e^{nv} K_{\mathcal{M}_2}.
\end{multline}

To remove the kinetic term for $v$ we perform a Weyl transformation $g_{\mu\nu} \mapsto g_{\mu\nu}^\text{W} = e^{(n-1)\, v(r)} g_{\mu\nu}$. Let us also introduce the dilaton field $\Phi(x) = e^{n \, v(x)}$
\begin{multline}\label{S-Wave:WeylTransformedAction}
    I_\text{W} =
    \frac{\lambda}{2} \int_{\mathcal{M}_2} d^2x \sqrt{- g_\text{W}} \Phi \left(
        R_\text{W}
        + \frac{n (n - 1)}{r_+^2}\Big(\Phi^{-1 - \frac{1}{n}}-\frac{r_+^2}{r_Q^2}\Phi^{-3 + \frac{1}{n}}\Big)
    \right)
    \\ + \lambda
    \int_{\partial\mathcal{M}} du \sqrt{-h_\text{W}} \Phi K_\text{W}.
\end{multline}
Previously we used the normalized tangent vector, that gives the unit induced metric $h^{\partial\mathcal{M}_2}_{uu} = \mathtt{t}^\mu \mathtt{t}^\nu g_{\mu\nu} = -1$ with $h_{\partial\mathcal{M}_2} = - 1$. Weyl transformed version $\sqrt{-h_\text{W}} = e^{-\frac{n-1}{2}v(r)}$ is not unit.

The action above is the action of 2D dilaton gravity with the potential,
\begin{equation}
    U(\Phi) =\frac{n (n - 1)}{r_+^2}\Big(\Phi^{-1 - \frac{1}{n}}-\frac{r_+^2}{r_Q^2}\Phi^{-3 + \frac{1}{n}}\Big).
\end{equation}
Its zero $U(\Phi_0) = 0$ corresponds to the trivial flat solution of the two-dimensional gravity
\begin{equation}
    \Phi_0 = 1 + O\Big(\frac{\tilde{\epsilon}^2}{n^3}\Big).
\end{equation}
Linear expansion near this point with the small correction to the dilaton $\Phi = \Phi_0 + \phi$
\begin{equation}
    U(\Phi) \simeq \frac{n(n-1)}{r_+^2} \left(2 - \frac{3}{n}\right) \phi
\end{equation}
results in JT gravity with the action
\begin{multline}
    I_\text{W} =
    \lambda\Phi_0 
    \left(\frac12 \int_{\mathcal{M}_2} d^2x \sqrt{- g_\text{W}} R_\text{W}
    + \int_{\partial\mathcal{M}_2} du \sqrt{-h_\text{W}} K_\text{W}\right)
    \\
    + \lambda
    \int_{\mathcal{M}_2} d^2x \sqrt{- g_\text{W}} \phi \left( R_\text{W} + \Big(2 - \frac{3}{n}\Big)\frac{n (n - 1)}{r_+^2} \right)
    + \lambda
    \int_{\partial\mathcal{M}_2} du \sqrt{-h_\text{W}} \phi K_\text{W}.
\end{multline}
The first bracket contains the Euler characteristic. It does not matter classically however plays a crucial role in two-dimensional quantum gravity controlling the topological series expansion. The rest of the action is JT gravity. The key property is the constant curvature of the bulk coming from the dilaton EoM:
\begin{equation}
     R_\text{W} = - \Big(2 - \frac{3}{n}\Big)\frac{n (n - 1)}{r_+^2}.
\end{equation}
In two-dimensional space this fixes the spacetime to be a patch of the AdS${}_2$ spacetime with the radius $L_2^2 = \frac{r_+}{n(n-1)(1-\tfrac{3}{2n})}$, i.e \textit{locally} the metric always can be written as,
\begin{equation}
    ds^2_\text{W} = \frac{L_2^2}{z^2} \left(- dt^2 + dz^2\right)
\end{equation}
The metric variation gives the following EoM,
\begin{equation}
    \left(\nabla_\mu\nabla_\nu - g_{\mu\nu} \nabla^2_{\mathcal{M}_2} + \frac{g_{\mu\nu}}{L_2^2}\right) \phi = 0.
\end{equation}
 and the corresponding solution within Poincare patch is the following (see~\cite{Maldacena:2016upp,Banerjee:2021vjy} for Euclidean version),
\begin{equation}
    \phi(t,z) = \frac{\alpha_1 + \alpha_2 t + \alpha_3 (z^2 - t^2)}{z},
\end{equation}
where $\alpha_i$ are real constants. We can rotate this vector of constants by using AdS${}_2$ isometries. For our case of the finite temperature blackhole we expect that the patch accessible from the far zone must be a Rindler patch \eqref{RindlerPatchTransform}. From \eqref{RindlerTransformObservation} we know that the solution,
\begin{equation}
\phi = \alpha \frac{2\tilde{\epsilon} + t}{z}
\end{equation}
is a static dilaton configuration in the Rindler patch.

The increase of the dilaton fluctuation with approach to the boundary of AdS${}_2$ corresponds to the expansion of $S_n$ with increase of $r$. When dilaton becomes large the JT approximation to the potential ceases to be valid and this corresponds to the entry to the transition region between near-horizon and far zone. To understand this more clearly let us consider a tortoise coordinate in the Reisner-Nordstr\"{o}m solution,
\begin{equation}
r_\ast = \int \frac{dr}{f(r)},\quad f(r) = \Bigg(1-\Big(\frac{r_+}{r}\Big)^{n-1}\Bigg)\Bigg(1-(1-\epsilon)\Big(\frac{r_+}{r}\Big)^{n-1}\Bigg)
\end{equation}

For RN solution $\Phi=\Big(\frac{r}{r_+}\Big)^n$ then in the limit $n\gg 1$, $\epsilon n=\mathrm{fix}$,
\begin{equation}
r_\ast = \frac{r_+}{n}\int \frac{d\Phi}{\Big(\Phi^{1-1/n} - 1\Big)\Big(\Phi^{1-1/n}-(1-\epsilon)\Big)}\simeq r_+ - \frac{r_+}{\epsilon n}\ln\Big(1+\frac{\epsilon n}{n(\Phi -1)}\Big)
\end{equation}
with integration constant chosen in such a way that the far zone $r>r_+$ was corresponding to $r_\ast > r$. It is easy to see that in the far zone $\Phi\sim n\rightarrow +\infty$. Then for the case $r_\ast < r_+$ we get,
\begin{equation}
1+\frac{1}{n}\frac{\epsilon n} {\Phi-1} = \exp\Big(n\epsilon\frac{r_+-r_\ast}{r_+}\Big)
\end{equation}

Thus the near-zone finite $r_\ast < r_+$ corresponds to $\phi=\Phi-1 \sim \frac{1}{n}$, whereas the whole range of finite $\phi$ correspond to $r_\ast \simeq r_+$. Thus in terms of the tortoise coordinates, the whole spacetime is split into two parts - the near-horizon region where the dilaton is $\sim\frac{1}{n}$ described by JT model and the far zone where the gravity of the black hole is miniscule.

%% file: TensorCorrections.tex
\section{Tensor corrections beyond s-wave approximation}\label{Section:Corrections}
Let us now introduce the small corrections $\mathpzc{h}_{MN}$ to the background solution $g_{MN}$ ~\cite{Flanagan:2005yc}.
\begin{equation}
    g_{MN} \mapsto g_{MN} + \frac{1}{\sqrt{2}M_{Pl}^{n/2}}\mathpzc{h}_{MN}.
\end{equation}
For the rest of the paper we will assume that $g_{MN}$ corresponds to the Reissner-Nordstr\"{o}m background.

As we are interested in the interaction of the black hole with the modes propagating from and toward infinity, we will restrict our attention to the transverse-traceless tensor modes
\begin{equation}
    \mathpzc{h}^{MN} = \mathpzc{t}^{MN}, \quad \mathpzc{t}^{\mu N} = 0, \quad
    \mathpzc{t}^{i}{}_{i} = 0, \quad
    \nabla_i^{(S^n)} \mathpzc{t}^{ij} = 0,
\end{equation}
with index $i$ corresponding to the angles $\theta^i$ on $S_n$.

Taking into account that the background is the equation of motion, the second variation of the bulk action reduces to,
\begin{equation}
    I^{(2)}_{\mathcal{M}}[\mathpzc h] = \int d^{(n+2)}x \sqrt{-g_{\mathcal{M}_D}} \left(
        \mathpzc h^{AB} \cdot \frac{\delta G_{AB}}{\delta g_{CD}} \cdot \mathpzc h^{CD}
    \right),
\end{equation}
\begin{multline}
    \frac{\delta G_{MN}}{\delta g_{AB}} \mathpzc{h}_{AB}= \tfrac12 \Big(
        - \mathpzc{h}_{MN} R
        + g_{MN} \mathpzc{h}^{AB} R_{AB}
        - \nabla^2 \mathpzc{h}_{MN}
        + \nabla_A \nabla_M \mathpzc{h}_N{}^A
        + \nabla_A\nabla_N\mathpzc{h}_M{}^A
        \\
        - g_{MN} \nabla_B\nabla_A \mathpzc{h}^{AB}
        + g_{MN} \nabla^2 \mathpzc{h}_A{}^A
        - \nabla_N\nabla_M\mathpzc{h}_A{}^A
    \Big)
\end{multline}
This term provides EoM on the corrections $\mathpzc{h}_{MN}$, that is discussed below.

The field can be expanded onto orthogonal tensor spherical modes $Y^{ij,\ell'\vec m'}(\vec \theta)$ with orbital moment $\ell$ and magnet numbers $\vec m$
\begin{equation}\label{Corrections:HarmonicsExpansion}
    \mathpzc{t}^{ij} = e^{-\frac{n}{2}v(r)} \sum_{\ell,\vec m} \mathfrak{t}_{\ell \vec m}(t,r) Y^{ij\ell'\vec m'}(\vec \theta),
\end{equation}
where the normalization $e^{-\frac{n}{2}v(r)}$ cancels additional multipliers in the quadratic action coming from $\sqrt{-g}$.
\begin{equation}
    \int_{S^n} d\Omega_n Y_{ij}^{\ell\vec m} Y^{ij\ell'\vec m'}
    = \delta_{\ell,\ell'} \delta_{\vec m, \vec m'}, \quad
    (\nabla^2_{(S^n)} + \lambda_{n,\ell}) Y_{ij}^{\ell\vec m} = 0, \quad
    \lambda_{n,\ell} = l(l + n - 1) - 2,
\end{equation}
for $l = 1, 2, \dots$~\cite{Kodama:2003jz}.

The two dimensional bulk action can be obtained from the compactification of the second correction
\begin{equation}
    I^{(2)}_{\mathcal{M}_2} = \sum_{\ell,\vec m} \int_{\mathcal{M}_2} d^2x \sqrt{-g_{\mathcal{M}_2}} \mathcal{L}_{\ell\vec m},
\end{equation}
where we introduce the notation for the Lagrangian for the correction with the moments $\ell$ and $\vec m$
\begin{equation}
    \mathcal{L}_{\ell\vec m} =
    - \mathfrak{t}_{\ell\vec m} \nabla^2_{\mathcal{M}_2} \mathfrak{t}_{\ell\vec m}
    + V(r) \mathfrak{t}_{\ell\vec m}^2, \quad
    V(r) = - (\nabla^2_{\mathcal{M}_2}e^{-\frac{n}{2}v}) - e^{-2v} \lambda_{n,\ell} - R
    + \mathcal{O}(n^1).
\end{equation}
The corresponding EoM $\delta\mathcal{L}_{\ell\vec m}/\delta \mathfrak{t}_{\ell\vec m} = 0$ can be solved within Fourier transformation
\begin{equation}\label{Corrections:FourierTransformations}
    \mathfrak t_{\ell\vec m}(t,r) = \int d\omega e^{i\omega t} q_{\ell\vec m}(r).
\end{equation}
Let us again employ the tortoise coordinates,
\begin{equation}
    dr_\ast = \frac{dr}{f(r)},
\end{equation}
noting that for $n\gg 1$ in the far zone $r>r_\ast$ the tortoise coordinate matches with the regular radial coordinate $r_\ast = r + \mathcal{O}(e^{-n})$, whereas in the near zone the whole range $r_\ast < r_+$ corresponds to a single value $r=r_+$.

The equation on the tensor modes then becomes a Schr\"{o}dinger-like equation,
\begin{equation}
    \omega^2 \mathfrak{t}_{\ell\vec m} + \partial_{\ast}^2 \mathfrak{t}_{\ell\vec m} - V_\text{Schr}(r_\ast) \mathfrak{t}_{\ell\vec m} = 0, \quad
    V_\text{Schr}(r)
    = 6 ff''
    + \frac{f'^2}{4}
    + \frac{3nff'}{2r}
    + \frac{f^2 n^2 r_+^2 \omega_\ell^2}{r^2}
    + \mathcal{O}(n^1)
\end{equation}
where $r_+ \omega_\ell = \frac{1}{2} + \frac{\ell}{n}$~\cite{Emparan:2014cia,Emparan:2014aba}. The potential $V_\text{Schr}(r)$. For the limit \eqref{LargeDLimit} this potential takes the from
\begin{equation}
    V_\text{Schr}(r_\ast)
    = \frac{n^2\omega_\ell^2 r_+^2}{r_\ast^2} \Theta(r_* - r_+)
    + \mathcal{O}(n^1),
\end{equation}

\begin{figure}
\centering
\includegraphics[width=0.8\textwidth]{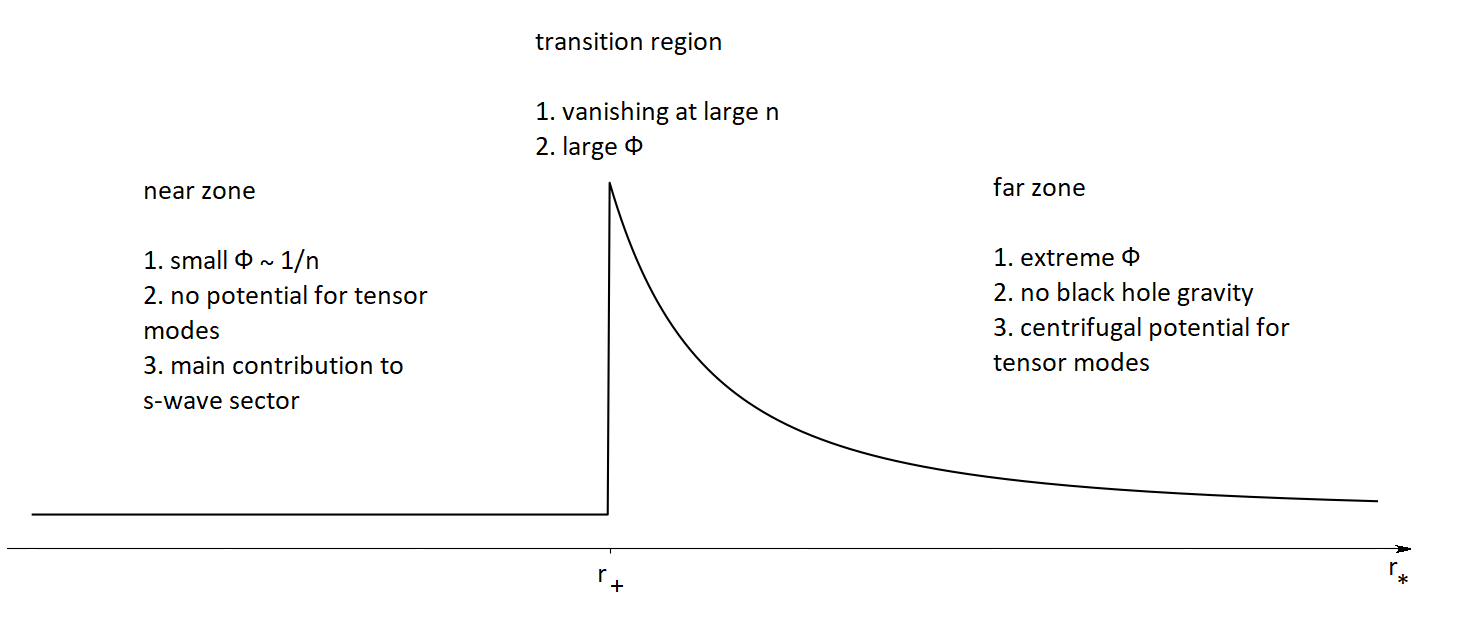}
\caption{The potential for the tensor modes at large $D$ and the location of the near and far zone}
\end{figure}

We allow only the waves that are ingoing with respect to the black hole horizon (i.e. we consider only the black hole, no white hole) Then the can be found as the plane waves in the near horizon region and Bessel function in far region
\begin{equation}%\label{Corrections:Solution}
    y^\text{near} = \mathpzc{T}_{n,\ell}(\omega) e^{i\omega r_*}, \quad
    y^\text{far}(r_+) =
    B_1 \sqrt{r} J_{\tfrac12\sqrt{1 + 4n^2r_+^2 \omega_\ell^2}}(r\omega)
    + B_2 \sqrt{r} Y_{\tfrac12\sqrt{1 + 4n^2r_+^2 \omega_\ell^2}}(r\omega).
    \label{BesselDecomposition}
\end{equation}

The transmit coefficient $\mathpzc{T}_{n,\ell}(\omega)$ can be found from the usual matching condition at the transition region (continuity of the solution and its derivatives). For $\omega \ll \frac{n}{r_+}$ the coefficient $B_2$ is expected to be of order of $\left|\mathpzc{T}_{n,\ell}(\omega)\right|^2$ and thus can be neglected.
\begin{equation}\label{Corresctions:TransitionRate}
    \mathpzc{T}_{n,\ell}(\omega) = e^{-i\omega r_+} \sqrt{r_+} J_{\tfrac12\sqrt{1 + 4n^2r_+^2 \omega_\ell^2}}(r_+ \omega)
    \sim \frac{e^{-i\omega r_+}}{\sqrt{\pi(n+2l)}}
    \left(\frac{e\omega r_+}{n+2l}\right)^\frac{n+2l}{2}
\end{equation}

The first order contribution to the boundary terms is,
\begin{equation}
    I_{\partial\mathcal{M}_D}^{(1)}[\mathpzc{t}]
    = \sqrt{2} M_{Pl}^{n/2} \int_{\partial\mathcal{M}_D} d^{n+1}y
    \left(\frac{\delta\sqrt{-h_{\partial\mathcal{M}_D}}}{\delta g^{ij}} K_{\mathcal{M}_D} + \sqrt{-h_{\partial\mathcal{M}_D}} \frac{\delta K_{\mathcal{M}_D}}{\delta g^{ij}}\right) \cdot \mathpzc{t}^{ij},
\end{equation}
where $\cdot$ is convolution.
Both the terms are $\propto g_{ij}$, that implies absence of first order corrections due to tracelessness of the perturbation $\mathpzc{t}_i{}^i = 0$.

The dominant contribution is thus happens at the second order,
\begin{multline}
    I_{\partial\mathcal{M}_D}^{(2)}[\mathpzc{t}] = 2\int_{\partial\mathcal{M}_D} d^{n+1}y \left(
        \mathpzc{t}^{ij} \cdot \frac{\delta^2\sqrt{-h_{\partial\mathcal{M}_D}}}{\delta g^{ij} \, \delta g^{kl}} \cdot \mathpzc{t}^{kl} K_{\mathcal{M}_D}
    \right.\\\left.
        + 2 \frac{\delta\sqrt{-h_{\partial\mathcal{M}_D}}}{\delta g^{ij}} \cdot \mathpzc{t}_{ij} \frac{\delta K_{\mathcal{M}_D}}{\delta g^{kl}} \cdot \mathpzc{t}^{kl}
        + \sqrt{-h_{\partial\mathcal{M}_D}} \mathpzc{t}^{ij} \cdot \frac{\delta^2 K_{\mathcal{M}_D}}{\delta g^{ij} \, \delta g^{kl}} \cdot \mathpzc{t}^{kl}
    \right).
\end{multline}

After taking into account the Weyl transformation of the two-dimensional metric \eqref{S-Wave:WeylTransformedAction} we can rewrite it as,
\begin{equation}
    I_\text{W}^{(2)}[\mathpzc{t}]
    = \frac{\text{Vol}S^n}{4} \int_{\partial\mathcal{M}_2} du \, 
    \sqrt{-h_\text{W}}
    \left(K_\text{W} + \frac{n + 1}{2}\mathtt{n}_\mu^\text{W} \nabla_\text{W}^\mu v\right)
    e^{nv} \mathpzc{K}
\end{equation}
where we have extracted volume of sphere for the proper comparison with s-wave terms and have introduced,
\begin{equation}
    \mathpzc{K} =
    \frac{r_+^n}{\text{Vol}S^n}\int d^n\theta \sqrt{\gamma}
        \mathpzc{t}_{ij} (t,\vec{x}) \,
        \mathpzc{t}_{kl} (t,\vec{\tilde{x}}).
\label{Kbeaut}
\end{equation}

%% file: Termalization.tex
\section{Canonical quantization in thermal bath}\label{Section:Quantization}

\subsection{Canonical quantization of the correlation function}

Canonical quantization of the perturbations on the solution in far region \eqref{BesselDecomposition} is the following
\begin{equation}
    \mathpzc{t}_{ij}^\text{far}(x^\mu) = \sum_{l, \vec{m}} \int_0^\infty d\omega \left(
        e^{i\omega t}\frac{\sqrt{r}}{\sqrt{2}r^{n/2}}
        J_{\nu}(\omega r) Y_{l\vec{m}, ij}(\theta) a_{l\vec{m}}^\dagger(\omega)
        + \text{h.c.}
    \right),
\end{equation}
where h.c. is Hermite conjugation. The raising $a_{l\vec{m}}^\dag(\omega)$ and lowering $a_{l\vec{m}}(\omega)$ operators are normalized
\begin{equation}
    [a_{l\vec{m}}(\omega), a_{\tilde{l}\vec{\tilde{m}}}^\dagger(\tilde{\omega})]
    = \delta_{l\vec{m},\tilde{l}\vec{\tilde{m}}} \delta(\omega-\tilde{\omega}).
\end{equation}
then the commutation relation for the fields,
\begin{align}
    [\mathpzc{t}_{ij}^\text{far}(\mathpzc{t},\vec{x}), \frac{d}{dt}t_{kl}^\text{far} (t,\vec{\tilde{x}})] =
    \sum_{l\vec{m}}\int_0^{+\infty} d\omega 
    \left(
        \frac{i\omega\sqrt{r\tilde{r}}}{2r^{n/2}\tilde{r}^{n/2}}
        J_{\nu}(\omega r) J_{\nu}(\omega \tilde{r})
        Y^\ast_{l\vec{m}, ij}(\theta) Y_{l\vec{m},kl}(\tilde{\theta})
        +\mathrm{h.c.}
    \right)
\end{align}
can be processed using the orthogonality relations on the spherical harmonics and Bessel functions
\begin{equation}
    \sum_{l\vec{m}} Y^\ast_{l\vec{m},ij}(\theta)Y_{l\vec{m},kl}(\tilde{\theta}) =
    \frac{\delta_{ij, kl}}{\sqrt{\gamma}}\delta^{(n)} (\theta-\tilde{\theta}), \quad
    \int_0^{+\infty} \omega r J_{\nu}(\omega r)J_{\nu}(\omega \tilde{r}) = \delta(r - \tilde{r})
\end{equation}
resulting in the canonical commutation relations
\begin{equation}
    [\mathpzc{t}_{ij}^\text{far} (t,\vec{x}), \frac{d}{dt}\mathpzc{t}_{kl}^\text{far} (t,\vec{\tilde{x}})]
    = i\frac{\delta_{ij, kl}}{r^n\sqrt{\gamma}} \delta^{(n)}(\theta-\tilde{\theta})
    \delta (r - \tilde{r})
    = i\delta_{ij, kl} \delta^{(n+1)} (\vec{x}-\vec{\tilde{x}})
\end{equation}

Let's replace in \eqref{Kbeaut} the quadratic form of the tensor fields by their avarage on the thermal state, i.e.
\begin{equation}
    \mathpzc{K}_\beta
    =
    \frac{r_+^n}{\text{Vol}S^n}\int d^n\theta \sqrt{\gamma}
    \left\langle:
        \mathpzc{t}_{ij}^\text{far} (t,\vec{x}) \,
        \mathpzc{t}_{kl}^\text{far} (t,\vec{\tilde{x}})
    : \right\rangle_\beta
\end{equation}
Substituting the decomposition at $r=r_+$
\begin{equation}
    \mathpzc{K}_\beta
    =
    \frac{r_+}{\text{Vol}S^n}\sum_{l\vec{m}}\int d^n\theta \sqrt{\gamma}\int_0^{+\infty} d\omega \Big(J_\nu (\omega r_+)\Big)^2 Y_{l\vec{m}, ab}(\theta)Y_{l\vec{m}, cd}(\theta)\langle :a^\dagger_{l\vec{m}}(\omega) a_{l\vec{m}}(\omega):\rangle_\beta 
\end{equation}
In the thermal state each mode behaves as an individual quantum harmonic oscillator in the thermal state. Therefore
\begin{equation}
    \mathpzc{K}_\beta
    =
    \frac{1}{\text{Vol}S^n}\frac{r_+}{\beta}
    \sum_{l\vec{m}}\int_0^{+\infty} d\omega \Big(J_\nu (\omega r)\Big)^2\cdot  \frac{\exp(-\beta\omega)}{1-\exp(\beta\omega)}
\end{equation}

\subsection{Asymptotic solution computation}

Using the asymptotics for the Bessel functions at large value of the argumentwe get,
\begin{equation}\label{Correlator1}
    \mathpzc{K}_\beta
    =
    \frac{1}{\text{Vol}S^n}\frac{r_+}{\beta}\sum_{l} d_l \int_0^{+\infty} d\omega
    \frac{1}{2\pi\nu\beta}\Big(\frac{e\omega r_+}{2\nu}\Big)^{2\nu}\frac{1}{e^{\beta\omega} - 1},
\end{equation}
where the spectral parameter and the degeneracy are,
\begin{equation}
    2\nu = \sqrt{1 + (n + 2\ell)^2}, \quad
    d_\ell = (n + 2\ell)\frac{(n - 2 + \ell)!}{(n-1)!\ell !}
\end{equation}

Introduce $w=\beta\omega$ then we can write \eqref{Correlator1} as,
\begin{multline}
    \mathpzc{K}_\beta
    =
    \frac{r_+^{-1}}{\text{Vol}S^n}\Big(\frac{r_+}{\beta}\Big)^2\sum_{l} d_l \frac{1}{2\pi\nu}\Big(\frac{e\omega r_+}{2\nu\beta}\Big)^{2\nu}\int_0^{+\infty} dw
    \frac{w^{2\nu}}{e^{w} - 1}=\\
    \frac{r_+^{-1}}{\text{Vol}S^n}\Big(\frac{r_+}{\beta}\Big)^2\sum_\ell\frac{d_\ell}{2\pi\nu}
    \Big(\frac{e r_+}{2\nu\beta}\Big)^{2\nu}
    \Gamma(2\nu + 1) (1-2^{-\nu}) \zeta(2\nu + 1)
\end{multline}
where $\zeta$ is the Riemann zeta function that for the large real argument $\nu \gg 1$ is $\zeta(2\nu + 1) \sim 1$. Using the Stirling approximation for the gamma function $\Gamma(2\nu + 1) \sim \sqrt{4\pi \nu}(2\nu/e)^{2\nu}$ we get,
\begin{equation}\label{Correlator2}
    \mathpzc{K}_\beta
    =
    \frac{r_+^{-1}}{\text{Vol}S^n}\Big(\frac{r_+}{\beta}\Big)^2\sum_{l}
    \frac{d_l}{\sqrt{\pi\nu}}
    \Big(\frac{r_+}{\beta}\Big)^{2\nu}
\end{equation}

The ratio $\xi = \frac{r_+}{\beta} = Tr_+ = \frac{n\epsilon}{4\pi}$ is assumed to be finite in our limit according to \eqref{NearExtremal:NearExtremalTemperature}. The sum is divergent with $\xi \geq 1$. The most probable cause for this is that in that case the thermal distribution
for the modes with $\omega \sim \frac{n}{r_+}$ becomes non-negligible. Therefore our approximation of the negligible $B_2$ coefficient in \eqref{BesselDecomposition} is not valid. For this reason we will restrict ourselves to the case $\xi < 1$. Then this sum is dominated by the values of $\ell$ that are large but still can be considered small compared to $n$. Then we can approximate the degeneracy with,
\begin{equation}
    d_\ell \simeq \frac{1}{2\pi \ell}\Big(\frac{ne}{\ell}\Big)^\ell
\end{equation}
Let us convert sum over the orbital momentum $\ell$ into the integral and use a new variable $s = \frac{\ell}{n}$. Note that,
\begin{equation}
    2\nu \simeq n + 2n s
\end{equation}
With the second term being important only inside the exponent. Then \eqref{Correlator2} is converted into
\begin{equation}
    \mathpzc{K}_\beta =
    \frac{r_+^{-1}}{\text{Vol}S^n}
    \frac{\xi^{n+2}}{2\pi^{3/2} n^{3/2}}
    \int_0^{+\infty} ds
    \frac{e^{ns}s^{-ns - 1}\xi^{2ns}}{\sqrt{1 + 2s}}
\end{equation}
The integral can be estimated within $n \gg 1$, where the integrand peaks at $s_\xi = \xi^2 + \mathcal{O}(1/n)$.
Thus we can estimate this contribution as,
\begin{equation}
    \mathpzc{K}_\beta =
    \frac{r_+^{-1}}{\text{Vol}S^n}
    \frac{1}{2\pi^{3/2} n^{3/2}}
    \frac{e^{n\xi^2}\xi^{n + 2}}{\sqrt{1 + 2\xi^2}}
\end{equation}
Use the asymptotic for the unit sphere area $\text{Vol}S^n \sim \sqrt{\tfrac{n}{\pi }} \left(\tfrac{2 \pi  e}{n}\right)^{n/2}$,
\begin{equation}
    \mathpzc{K}_\beta =
    r_+^{-1}
    \sqrt{\frac{n}{\pi }} \left(\frac{2 \pi  e}{n}\right)^{-n/2}
    \frac{1}{2\pi^{3/2} n^{3/2}}
    \frac{e^{n\xi^2}\xi^{n + 2}}{\sqrt{1 + 2\xi^2}}.
\end{equation}

For large $n$ the correlator $\mathpzc{K}_\beta \rightarrow +\infty$. This is offset by this term not being multiplied by $(M_{Pl}r_+)^n$ unlike s-wave terms. The relation between this two large factors determines the strength of the backreaction of the quantum fluctuations on the background.

%% file: Conclusion.tex
\section*{Conclusion}

In this paper we have used a combination of the near-extremality and large dimension number limits to keep the temperature of the black hole finite. In this case we get the Jackiw-Teitelboim gravity in the near-horizon zone that is expected to be at the Hawking temperature of the black hole. Gravity in the near zone can be considered classical if $\lambda\gg 1$. However, extra factor $\text{Vol} S^n$ that goes to zero when $n\rightarrow +\infty$ means that the gravity may become quantum in the near zone even if the modes with $E\sim \frac{1}{r_+}\ll M_{Pl}$ remain to be weakly coupled in the far zone.

The interaction of the non-s-wave modes with this region becomes very weak. However, unlike in case of the finite dimension extremal limit, from the point of view of the ingoing-outgoing radiation the transition region becomes vanishingly small and thus it seems that the system is separated into two parts:
\begin{itemize}
\item The free field theory in the Minkowsky spacetime of the far zone
\item The two-dimensional JT gravity of the near-horizon zone with boundary at $\Phi\gg \frac{1}{n}$ representing a transition region, where the sources produced by the interaction with the free waves are introduced.
\end{itemize}

Thus, such limit may present a very good setup to study the quantum effects of the black hole. The important thing about our consideration is that we have demonstrated that despite vanishingly small couplings of each individual mode to the near-horizon region, their collective impact may actually be significant and may actually deform a near-horizon zone model.
This is actually expected as the finite temperature also mean a significant impact of the Hawking radiation on the black hole dynamics.